\documentclass[aps,prl,twocolumn,superscriptaddress,showpacs,10pt]{revtex4-1}

\usepackage{xcolor,graphicx,ulem}
\usepackage{amsmath,amssymb}
\usepackage[colorlinks=true,urlcolor=blue,citecolor=blue,linkcolor=blue]{hyperref}

\begin{document}

\title{Quantum Correlations from the Conditional Statistics of Incomplete Data}

\author{J. Sperling}\email{jan.sperling@physics.ox.ac.uk}
	\affiliation{Arbeitsgruppe Theoretische Quantenoptik, Institut f\"ur Physik, Universit\"at Rostock, D-18051 Rostock, Germany}
	\affiliation{Clarendon Laboratory, University of Oxford, Parks Road, Oxford OX1 3PU, United Kingdom}
\author{T. J. Bartley}
	\affiliation{Clarendon Laboratory, University of Oxford, Parks Road, Oxford OX1 3PU, United Kingdom}
	\affiliation{Universit\"at Paderborn, Warburger Strasse 100, 33098 Paderborn, Germany}
\author{G. Donati}
	\affiliation{Clarendon Laboratory, University of Oxford, Parks Road, Oxford OX1 3PU, United Kingdom}
\author{M. Barbieri}
	\affiliation{Clarendon Laboratory, University of Oxford, Parks Road, Oxford OX1 3PU, United Kingdom}
	\affiliation{Dipartimento di Scienze, Universit\`a degli Studi Roma Tre, Via della Vasca Navale 84, 00146 Rome, Italy}
\author{X.-M. Jin}
	\affiliation{Clarendon Laboratory, University of Oxford, Parks Road, Oxford OX1 3PU, United Kingdom}
	\affiliation{State Key Laboratory of Advanced Optical Communication Systems and Networks, Institute of Natural Sciences \& Department of Physics and Astronomy, Shanghai Jiao Tong University, Shanghai 200240, China}
\author{A. Datta}
	\affiliation{Department of Physics, University of Warwick, Coventry CV4 7AL, United Kingdom}
\author{W. Vogel}
	\affiliation{Arbeitsgruppe Theoretische Quantenoptik, Institut f\"ur Physik, Universit\"at Rostock, D-18051 Rostock, Germany}
\author{I. A. Walmsley}
	\affiliation{Clarendon Laboratory, University of Oxford, Parks Road, Oxford OX1 3PU, United Kingdom}

\date{\today}

\begin{abstract}
	We study, in theory and experiment, the quantum properties of correlated light fields measured with click-counting detectors providing incomplete information on the photon statistics.
	We establish a correlation parameter for the conditional statistics, and we derive the corresponding nonclassicality criteria for detecting conditional quantum correlations.
	Classical bounds for Pearson's correlation parameter are formulated that allow us, once they are violated, to determine nonclassical correlations via the joint statistics.
	On the one hand, we demonstrate nonclassical correlations in terms of the joint click statistics of light produced by a parametric down conversion source.
	On the other hand, we verify quantum correlations of a heralded, split single-photon state via the conditional click statistics together with a generalization to higher-order moments.
	We discuss the performance of the presented nonclassicality criteria to successfully discern joint and conditional quantum correlations.
	Remarkably, our results are obtained without making any assumptions on the response function, quantum efficiency, and dark-count rate of the photodetectors.
\end{abstract}

\maketitle

\paragraph*{Introduction.--}
	The photon statistics of light lie at the heart of quantum optics.
	Starting with the landmark experiments of Hanbury Brown and Twiss~\cite{HBT56}, the photon distribution can be used to determine whether or not the field is consistent with a classical description.
	The classical theory of radiation is satisfied when the quantum system when a state of light can be written as classical mixtures of coherent states.
	Whenever such a description fails, this is referred to as nonclassical light~\cite{TG65}.
	However, uncovering nonclassical phenomena in the presence of loss, decoherence, and using imperfect devices is typically very challenging~\cite{CGLM14,PADLA10,DZTDSW11}.
	Moreover, quantum behavior defines the foundation of modern quantum technologies~\cite{KMNRDM07}, which results in a pressing need for reliable tools to process quantum states that are robust against real world environments.
	As the physical implementations become ever more complex, it is helpful if analysis techniques place minimal assumptions on the underlying principle of operation of such devices.

	Characterizing the correlations of two beams of light, $A$ and $B$, can be done in two ways.
	First, one can access the nonclassical character of light via joint correlation functions, similar to the proposal of Hanbury Brown and Twiss~\cite{HBT56}.
	Second, one can ask: ``How well is the outcome of a measurement in system $B$ determined for a fixed outcome in $A$?'' and ``Is the degree of determination compatible with classical light?''
	While the first approach is based on the joint probability distribution, the latter questions address the conditional statistics.
	In quantum systems, the conditional type of correlations leads to quantum effects such as steering~\cite{WJD07,CPMA14}.
	However, conditional correlations are typically not studied in the context of nonclassical radiation fields.

	In general, quantum features of the photon number statistics can be accessed with nonclassicality tests~\cite{M79,AT92,KDM77,V08,MBWLN10,ALCS10,AOB12}.
	However, detectors that directly measure the photon distribution are not commercially available as they require, for instance, cryogenic cooling; see~\cite{BC10} for an overview.
	To gain significant albeit incomplete information about a given state of light, it is possible to consider technically much simpler systems consisting of multiple on-off detectors, that is, avalanche photodiodes (APDs) in the Geiger mode~\cite{ZABGGBRP05,BGGMPTPOP11}.
	Examples of such schemes are given by CCD detectors~\cite{WDSBY04,HPHP05,LBFD08,BDFL08,CWB14} and multiplexing layouts~\cite{RHHPH03,ASSBW03,FJPF03,LFCPSREPW08}.
	In the latter scenario, one splits light into several spatial beams or temporal bins with smaller intensities, each being measured with an APD.
	The main feature of the resulting click-counting statistics is its binomial character~\cite{SVA12}, which significantly differs from the Poissonian form of the photon-number distribution; see also Refs.~\cite{R14,MSB16}.
	Therefore, the nonclassicality probes have to be adjusted properly to correctly uncover nonclassical light~\cite{SVA12a,BDJDBW13}.
	Such a technique directly identifies quantumness in integrated waveguides~\cite{HSPGHNVS16} or systems with high losses~\cite{SBVHBAS15}.
	These state-of-the-art implementations underline the functionality of click-counting detectors for applications in realistic scenarios.
	However, such techniques are specific to joint correlations of quantum states, and they may fail to uncover the conditional nonclassicality.

	In this Letter, we present a generalized approach to handling statistics from multimode states which place minimal assumptions on the detectors.
	For this reason, we formulate a conditional correlation parameter and derive its bounds for classical light.
	We additionally compute the classical limits for Pearson's correlation coefficient for inferring nonclassicality of the joint click distribution.
	By implementing a parametric-down-conversion source, we produce differently quantum correlated states of light that are probed with our approaches.
	Our method for the nonclassical conditional statistics is generalized and applied to access higher-order moments.

	The primary aim of this work is to demonstrate how a general theoretical analysis of click statistics can be used to highlight nonclassical behavior from a range of quantum states.
	We illustrate the utility of this analysis with several examples from typical experimental data.
	The strength of the methodology that will be introduced lies in its ability to identify and discriminate joint and conditional nonclassicality with minimal assumptions about how the data were acquired and the characteristics of the detector.
	In particular, this nonclassicality can be correctly determined even when the raw data arising from different states appear very similar, due to the effects of noise and loss.

\paragraph*{Click-counting detectors.--}
	Let us briefly recall the theory of click-counting devices.
	The probability that $a=0,\ldots,N_A$ out of $N_A$ APDs produce a coincidence click event is~\cite{SVA12}
	\begin{align}\label{eq:ClickStat}
		c(a)=\left\langle{:}
			\binom{N_A}{a}\hat\pi_A^{a}(\hat 1-\hat\pi_A)^{N_A-a}
		{:}\right\rangle,
	\end{align}
	with $\hat\pi_A=\hat 1-\exp(-\Gamma(\hat n_A/N_A))$.
	In this normally ordered expectation value, $\langle{:}\cdots{:}\rangle$, $\hat n_A$ is the photon number operator and $\Gamma$ is the so-called detector response function.
	One typically considers a linear response $\Gamma(x)=\eta x+\nu$ (quantum efficiency $\eta$; dark count rate $\nu$).
	Here, we do not assume any form of $\Gamma$.
	Moreover, the click-statistics gives incomplete information about the photon statistics, because an inversion of a finite number of clicks to an infinite number of possible photon states is, in principle, impossible.
	Similar to the single mode case in Eq.~\eqref{eq:ClickStat}, one can describe the joint click-counting distribution $c(a,b)$ for two light fields $A$ and $B$~\cite{SVA13}.
	In this case, the joint statistics is given by the operators $\hat \pi_A$ and $\hat\pi_B$ as well as the numbers of APDs $N_{A}$ and $N_B$.

	The features of the click-counting statistics and the quantum statistics are closely related, for example,
	\begin{align}\label{eq:momentREL}
		\langle{:}(\Delta\hat\pi_A)^2{:}\rangle {=} \frac{ N_A{\rm Var}_{c(a)}(a) {-} {\rm E}_{c(a)}(a) \big(N_A {-} {\rm E}_{c(a)}(a)\big) }{N_A^2(N_A {-} 1)} {,}
	\end{align}
	with $\Delta\hat\pi_A=\hat\pi_A-\langle{:}\hat\pi_A{:}\rangle$ and the symbol ${\rm E}_{c(a)}$ and ${\rm Var}_{c(a)}$ stand for the expectation value and the variance of the marginal statistics $c(a)=\sum_{b=0}^{N_B} c(a,b)$, respectively~\cite{SVA12a,SVA13}.
	It has been similarly shown for the covariance that
	\begin{align}
		{\rm Cov}_{c(a,b)}(a,b)=N_AN_B\langle{:}\Delta\hat\pi_A\Delta\hat\pi_B{:}\rangle.
	\end{align}

	For measuring quantum effects of a single mode, one can define the binomial $Q$ parameter~\cite{SVA12a},
	\begin{align}\label{eq:Qbinomial}
		Q_{c(a)}=&\frac{N_A{\rm Var}_{c(a)}(a)}{{\rm E}_{c(a)}(a)\big(N_A-{\rm E}_{c(a)}(a)\big)}-1\stackrel{\rm cl.}{\geq}0.
	\end{align}
	For classical light, this parameter is non-negative.
	If this condition is violated, one has {\it sub-binomial} light~\cite{BDJDBW13,HSPGHNVS16}.
	However, the value of the binomial $Q$ parameter cannot give information about the correlations between $a$ and $b$.

\paragraph*{Experimental setup and generated states.--}
	\begin{figure}[h]
		\includegraphics[width=8cm]{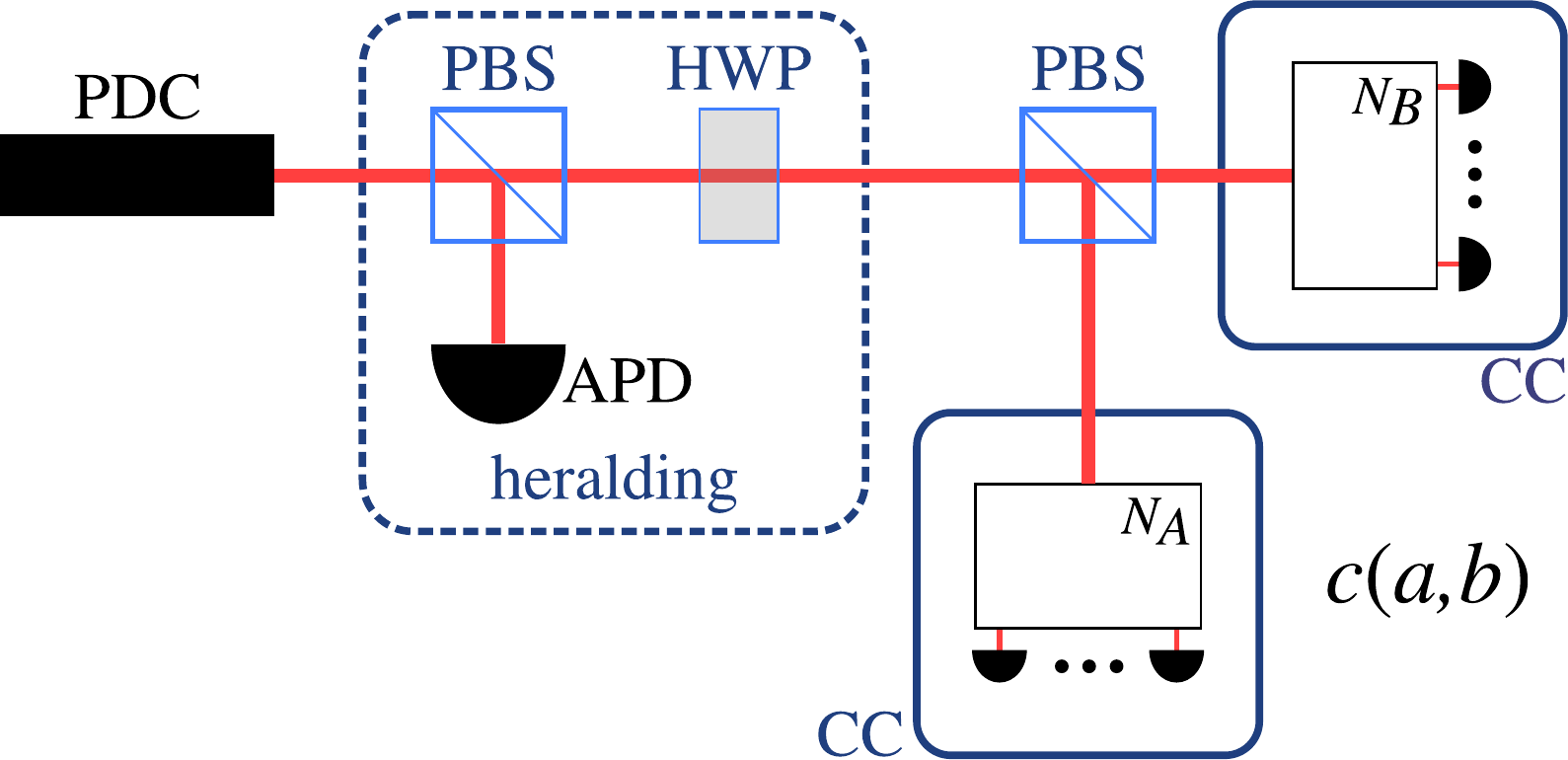}
		\caption{(Color online)
			Schematic of the generation of a two-mode squeezed-vacuum state (excluding the dashed framed pattern) with a parametric down-conversion (PDC) source and a single photon split at a polarizing beam splitter (PBS).
			The latter one is generated by heralding onto the click of the single APD and rotated in polarization with a half-wave plate (HWP).
			Each resulting mode is sent to an 8-bin time-multiplexed detector serving as our click counter (CC).
		}\label{fig:Setup}
	\end{figure}

	We illustrate the measurement layout in Fig.~\ref{fig:Setup}.
	In our experiment, we adopt time-multiplexed click-counting detectors, which separate the incoming light into $N_A=N_B=8$ distinct bins, by the use of two cascaded, unbalanced fiber Mach-Zehnder interferometers~\cite{ASSBW03,BDJDBW13}.
	To generate our two-mode squeezed vacuum (TMSV) states, $(1{-}\lambda^2)^{1/2}\sum_{n=0}^\infty \lambda^n |n\rangle_A|n\rangle_B$ ($0{<}\lambda{<}1$), we pump a nonlinear KDP crystal via type-II collinear parametric down conversion~\cite{MLSWUSW08}.
	This produces photon pairs in orthogonal polarization modes.
	These modes are split at a polarizing beam splitter and directed to two spatially separated time-multiplexed detectors.
	We obtain a set of joint counts $C(a,b)$, which we normalize to obtain the joint click probabilities $c(a,b)$.
	At a pump pulse repetition rate of $250\,{\rm kHz}$, we obtain single-click count rates of the order $\sim1\,{\rm kHz}$, and taking data for approximately $10$ minutes yields $\sim10^7$ data points.
	The split-photon (SP) states, $t|1\rangle_A|0\rangle_B+(1{-}t^2)^{1/2}|0\rangle_A|1\rangle_B$ ($0{<}t{<}1$), are produced by heralding a single photon from the parametric down-conversion process, and splitting it into two modes.
	The analysis is carried out in the same fashion as for the TMSV state, but with $\sim10^{6}$ data points.
	Heralding decreases the data rate and, therefore, the overall counts during stable operation of the experiment.
	For technical details on the error analysis and the experiment, we also refer to the Supplemental Material~\cite{SM} (Sec.~D and Sec.~E).

	A single case per state would be sufficient to illustrate the utility of our analysis.
	However, we also show how this technique captures the variability under standard experimental conditions;
	in this respect, we analyzed statistics arising from two TMSV states and three SP states.
	We have click numbers in the interval $0.03\leq {\rm E}_{c(a,b)}(a{+}b)\leq0.11$ for the states TMSV$_1$ and TMSV$_2$, as well as for the three states SP$_{1,2,3}$ (cf. Table~\ref{tab:HONcl} and Sec.~E in~\cite{SM}).
	As a classical reference, we also characterized a two-mode coherent state, $|\alpha\rangle_A|\beta\rangle_B$, by blocking the signal and sending laser light into the free port of the last PBS in Fig.~\ref{fig:Setup}.

\paragraph*{Conditional quantum correlations.--}
	The conditional statistics, $p(b|a)=p(a,b)/p(a)$, determines how well the outcome of $b$ is determined for a given $a$ value.
	The variance of the conditional statistics, ${\rm Var}_{p(b|a)}(b)$, describes the uncertainty of $b$ for a fixed $a$.
	In addition, if the outcome of a condition $a$ is more likely than another one, this should also have a larger contribution to this uncertainty.
	Hence, we formulate a correlation measure in terms of the mean conditional variance ${\rm E}_{p(a)}\big({\rm Var}_{p(b|a)}(b)\big)$ in the form
	\begin{align}\label{eq:CondCorCoeff}
		\kappa_{p(b|a)}=&1-\frac{{\rm E}_{p(a)}\big({\rm Var}_{p(b|a)}(b)\big)}{{\rm Var}_{p(b)}(b)}.
	\end{align}
	In~\cite{SM} (Sec.~B), we characterize the {\it conditional correlation coefficient} in Eq.~\eqref{eq:CondCorCoeff}.
	There, we show that $0\leq\kappa_{p(b|a)}\leq1$ and that the lower and the upper bound is attained for any uncorrelated and any perfectly correlated probability distribution, respectively.

	For the conditional click statistics, let us formulate the bounds of $\kappa_{c(b|a)}$ for classical light.
	For a classical state, the conditional statistics $c(b|a)$ is also classical~\cite{SM} (Sec.~A).
	Thus, we find the following constraint for classical states:
	\begin{align}\label{eq:NclKappa}
		\kappa_{c(b|a)}\stackrel{\rm cl.}{\leq}\kappa_{c(b|a)}^{\rm cl.max},
	\end{align}
	with
	\begin{align}
		\kappa_{c(b|a)}^{\rm cl.max}=1-\frac{{\rm E}_{c(a)}\Big({\rm E}_{c(b|a)}(b)\big(N_B-{\rm E}_{c(b|a)}(b)\big)\Big)}{N_B{\rm Var}_{c(b)}(b)}.
	\end{align}
	The latter bound has been obtained by inserting constraints on the conditional variance of a classical signal into Eq.~\eqref{eq:CondCorCoeff}.
	That is, the conditional binomial parameter $Q_{c(b|a)}$ implies in this case ${\rm Var}_{c(b|a)}(b)\geq {\rm E}_{c(b|a)}(b)(N_B-{\rm E}_{c(b|a)}(b))/N_B$, cf. Eq.~\eqref{eq:Qbinomial}.

	Whenever inequality~\eqref{eq:NclKappa} is violated, the degree of determination of $b$ in terms of $\kappa_{c(b|a)}$ is too large to be compatible with classical light.
	Hence, we have constructed a measure for quantum correlations for conditional click-counting statistics.
	The given bound $\kappa_{c(b|a)}^{\rm cl.max}$ is tight, as for any binomial click statistics, for instance for coherent states, holds ${\rm Var}_{c(b|a)}(b)= {\rm E}_{c(b|a)}(b)(N_B-{\rm E}_{c(b|a)}(b))/N_B$.
	For comparison, we derived similar bounds for the photon-counting theory~\cite{SM} (Sec.~C).

\begin{figure}[ht]
		\includegraphics[width=8.5cm]{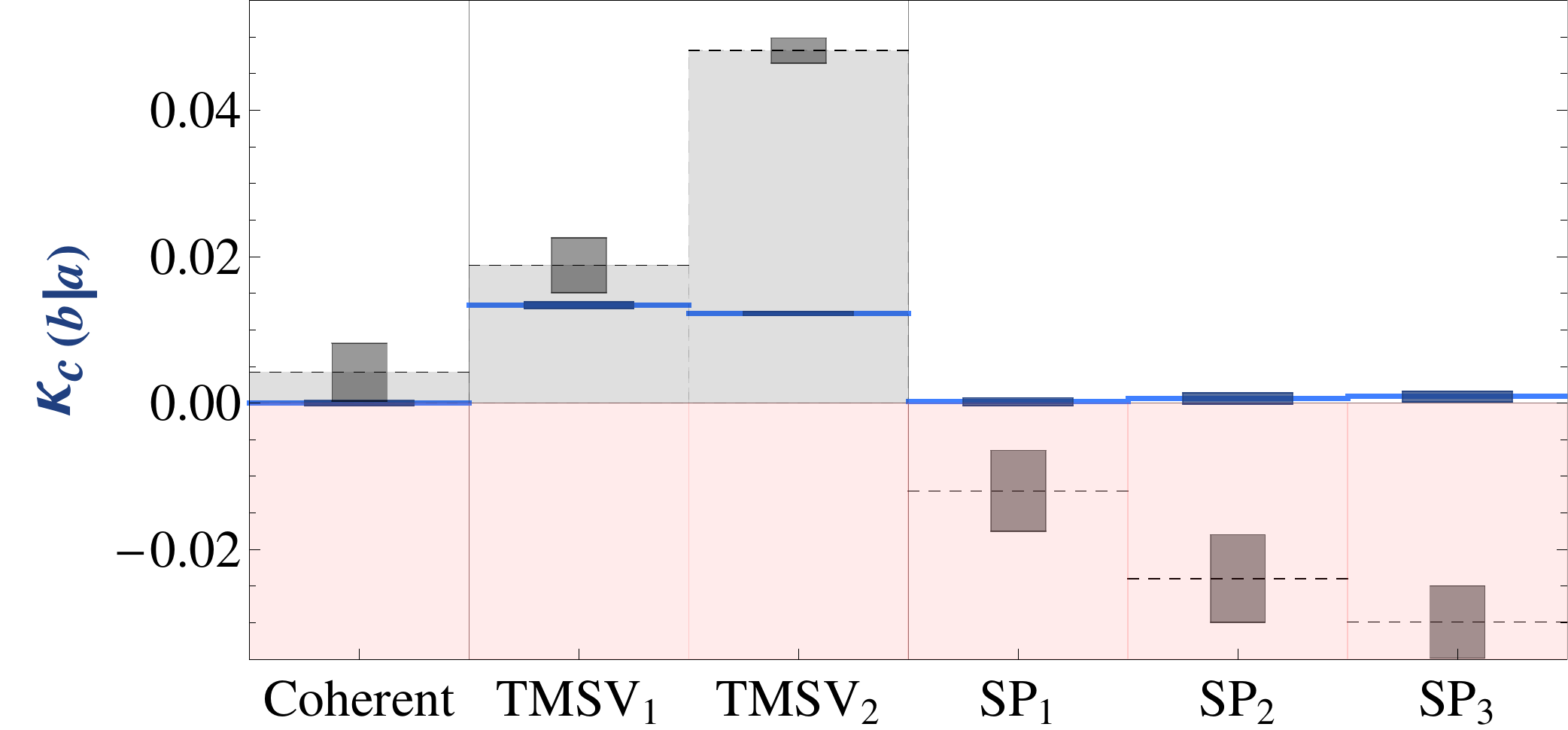}
		\caption{(Color online)
			Results of conditional correlation coefficient $\kappa_{c(b|a)}$ (blue, solid lines) and the corresponding classical bounds $\kappa_{c(b|a)}^{\rm cl.max}$ (black, dashed lines) are shown including the corresponding error bars.
			Note that the plotted linewidth of $\kappa_{c(b|a)}$ is larger than its error bar in some cases.
			The light gray areas show the classically allowed ranges, cf. inequality~\eqref{eq:NclKappa}.
			The (red) area $\kappa_{c(b|a)}{<}0$ corresponds to the unphysical values.
		}\label{fig:ConditionalCor}
	\end{figure}

	We directly applied the conditional correlation parameter to our measured data, see Fig.~\ref{fig:ConditionalCor}.
	The uncorrelated, classical coherent state is compatible with the expectation $\kappa_{c(b|a)}\approx\kappa_{c(b|a)}^{\rm cl.max}\approx0$.
	The quantum correlations of the TMSV state are not accessible with the conditional correlation parameter, $\kappa_{c(b|a)}\leq \kappa_{c(b|a)}^{\rm cl.max}$.
	For the SP states, we encounter the fact that $\kappa_{c(b|a)}^{\rm cl.max}<0$.
	Since $\kappa_{c(b|a)}$ is necessarily non-negative, we have, in such a case, $\kappa_{c(b|a)}^{\rm cl.max}<0\leq\kappa_{c(b|a)}$, which violates the classical constraint~\eqref{eq:NclKappa}.
	Thus, the SP states exhibit a nonclassical conditional correlation.

\paragraph*{Classical bounds for joint correlations.--}
	For an arbitrary joint probability distribution $p(a,b)$, a well-established measure of joint correlations is {\it Pearson's correlation coefficient}~\cite{P95}
	\begin{align}\label{eq:PCC}
		\gamma_{p(a,b)}=\frac{{\rm Cov}_{p(a,b)}(a,b)}{\sqrt{{\rm Var}_{p(a,b)}(a){\rm Var}_{p(a,b)}(b)}}.
	\end{align}
	For statistically independent random variables $a$ and $b$, we have $\gamma_{p(a,b)}(a,b)=0$.
	A (negative)positive value characterizes (anti-)correlations.
	The ultimate bound for any statistics is $|\gamma_{p(a,b)}(a,b)|\leq 1$.

	Let us derive the bound for classical states.
	The covariance of the joint click-counting distribution can be bounded for classical states via a normally ordered form of the Cauchy-Schwarz inequality
	\begin{align}
		\left|\langle{:}\Delta\hat\pi_A\Delta\hat\pi_B{:}\rangle\right|\stackrel{\rm cl.}{\leq}&
		\left|\langle{:}(\Delta\hat\pi_A)^2{:}\rangle\right|^{1/2}\left|\langle{:}(\Delta\hat\pi_B)^2{:}\rangle\right|^{1/2}.
	\end{align}
	The normally ordered variances can be given in terms of Eq.~\eqref{eq:momentREL} for systems $A$ and $B$.
	Using the definitions of $\gamma_{c(a,b)}$ in Eq.~\eqref{eq:PCC}, $Q_{c(a)}$, and $Q_{c(b)}$, we conclude
	\begin{align}\label{eq:PearsonCl}
		-\gamma_{c(a,b)}^{\rm cl.max.}\stackrel{\rm cl.}{\leq}\gamma_{c(a,b)}\stackrel{\rm cl.}{\leq}\gamma_{c(a,b)}^{\rm cl.max.},
	\end{align}
	where
	\begin{align}\label{eq:PearsonBound}
		\gamma_{c(a,b)}^{\rm cl.max}{=}\!\left|\frac{N_AN_BQ_{c(a)}Q_{c(b)}}{(N_A{-}1)(N_B{-}1)(Q_{c(a)}{+}1)(Q_{c(b)}{+}1)}\right|^{1/2}\!.
	\end{align}
	Interestingly, the bound for a classical Pearson's correlation coefficient $\gamma_{c(a,b)}^{\rm cl.max}$ can be written solely in terms of the measured binomial $Q$ parameters for $A$ and $B$, as well as the numbers of APDs, $N_A$ and $N_B$.
	In~\cite{SM} (Sec.~C), we give a similar relation for the photon-counting detectors in terms of the Mandel $Q$ parameter~\cite{M79}.
	There, we also construct a nonlinearly, perfectly correlated state whose quantum correlations cannot be inferred via $\gamma_{p(a,b)}$, but can be uncovered with $\kappa_{p(b|a)}$.

\begin{figure}[ht]
		\includegraphics[width=8.5cm]{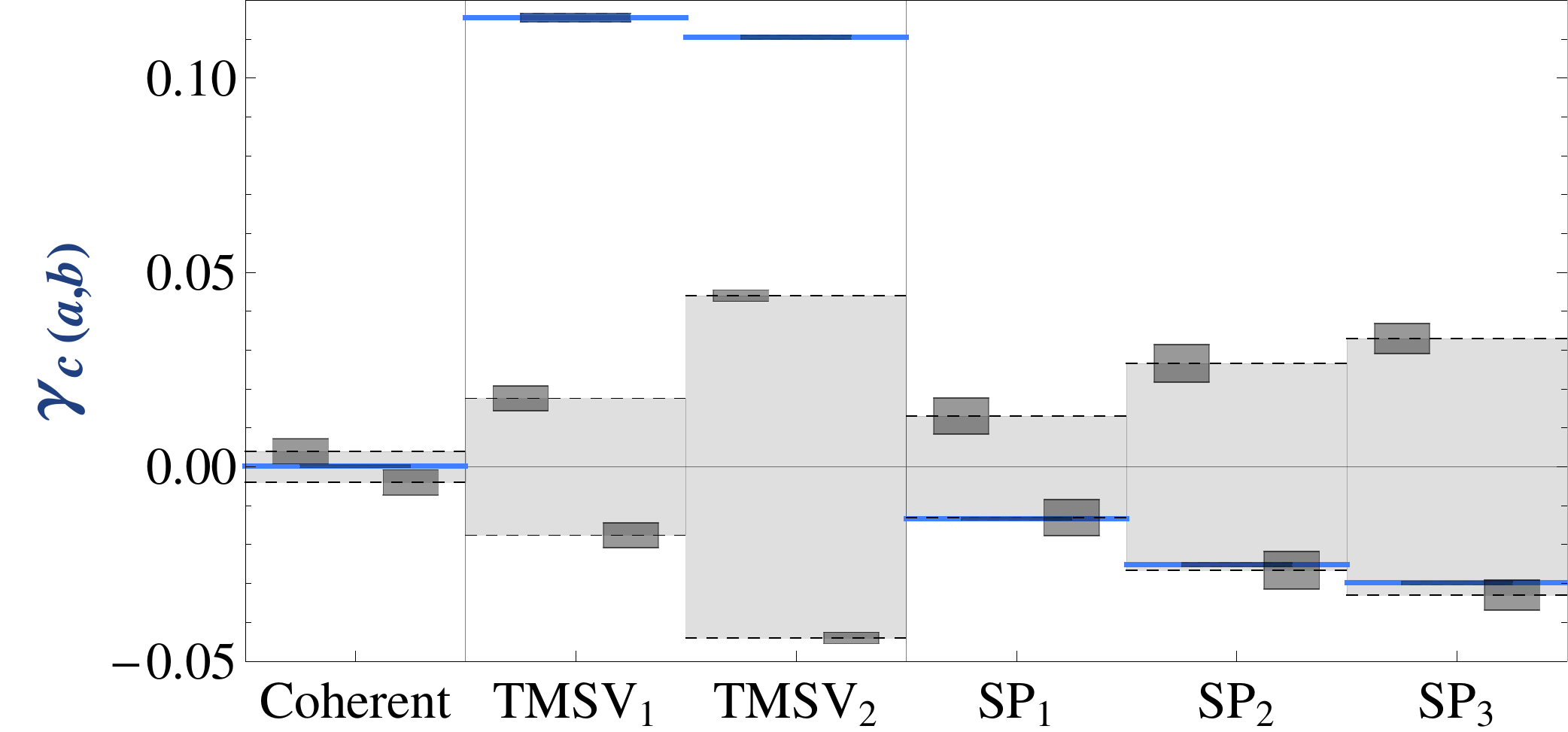}
		\caption{(Color online)
			Results of Pearson's correlation coefficient $\gamma_{c(a,b)}$ (blue, solid lines) and the corresponding classical bounds $\pm\gamma_{c(a,b)}^{\rm cl.max}$ (black, dashed lines) are shown.
			The light gray areas show the classically allowed ranges.
		}\label{fig:PearsonCor}
	\end{figure}

	In Fig.~\ref{fig:PearsonCor}, we show the application of the classical constraint~\eqref{eq:PearsonCl} to our measurements.
	The value of the coherent state is consistent with the expected value $\gamma_{c(a,b)}{=}0$.
	The TMSV and the SP states are significantly correlated $\gamma_{c(a,b)}{>}0$ and anticorrelated $\gamma_{c(a,b)}{<}0$, respectively.
	In addition, the TMSV states clearly exceed the classical bound $\gamma_{c(a,b)}^{\rm cl.max}$, while this is not true for the SP states.
	Thus, the TMSV states show joint nonclassical correlations.

	Our results show that the correlation parameters $\kappa_{c(b|a)}$ and $\gamma_{c(a,b)}$ are sensitive to different kinds of quantum correlations and complement each other.
	On the one hand, the conditional correlation coefficient detects quantum correlations of the SP states in terms of the conditioned click statistics $c(b|a)$.
	On the other hand, Pearson's correlation coefficient is sensitive to correlations of the joint click statistics $c(a,b)$ which applies to the TMSV states.

\paragraph*{Higher-order conditional correlations.--}
	The error bars in Fig.~\ref{fig:ConditionalCor} for the SP states indicate that the significance of verified conditional quantum correlations decreases with decreasing summed click number ${\rm E}_{c(a,b)}(a{+}b)$.
	Hence, we will extend our method to higher-order conditional correlations.
	The conditional statistics $c(b|a)$ can be written in a form similar to Eq.~\eqref{eq:ClickStat} in terms of conditional, normally ordered expectation values:
	$c(b|a)=\langle{:}\binom{N_B}{b}\hat\pi_{B}^b(\hat 1_A-\hat\pi_{B})^{N_B-b}{:}\rangle_{|a}$.
	In Refs.~\cite{SVA13,SBVHBAS15}, we introduced and applied a method to characterize higher-order nonclassicality based on normally ordered moments of the click statistics.
	Here, the corresponding condition can be rewritten as
	\begin{align}\label{eq:NclHO}
		\mathfrak N_{c(b|a)}=\min_{a=0,\dots,N_A}\left\{\langle{:}\hat f_a{\!}^\dagger\hat f_a{:}\rangle_{|a}\right\}\stackrel{\rm cl.}{\geq}0,
	\end{align}
	with $\hat f_a=\sum_{m=0}^{N_B/2} f_{m|a}\hat\pi_{B}^m$ and using the coefficient vector $(f_{0|a},\ldots,f_{N_B/2|a})^{T}$.
	The coefficient vectors are chosen to be the eigenvectors to the minimal eigenvalue of the conditional matrix of moments, $\big(\langle{:}\hat\pi_{B}^{m+m'}{:}\rangle_{|a}\big)_{m,m'=0}^{N_B/2}$, for minimizing the individual normally ordered, conditional expectation values $\langle{:}\hat f_a{\!}^\dagger\hat f_a{:}\rangle_{|a}$~\cite{SVA13,SBVHBAS15}.

	\begin{table*}
		\caption{
			Success of the nonclassicality correlation tests.
			The symbol ``$\boldsymbol\checkmark$'' describes a significant verification of quantum correlations, otherwise, we put ``$\boldsymbol\times$''.
			The the summed click number ${\rm E}_{c(a,b)}(a{+}b)$ and the higher-order conditional nonclassicality number $\mathfrak N_{c(b|a)}$ [Eq.~\eqref{eq:NclHO}] are explicitly given in the second and last row, respectively, including their relative errors.
		}\label{tab:HONcl}
	\begin{tabular}{c c c c c c}
		\hline\hline
		State & ${\rm E}_{c(a,b)}(a{+}b)$ & $\kappa_{c(b|a)}{>}\kappa^{\rm cl.max}_{c(b|a)}$ & $|\gamma_{c(a,b)}|{>}\gamma^{\rm cl.max}_{c(a,b)}$ & $\mathfrak N_{c(b|a)}{<}0$ & $\mathfrak N_{c(b|a)}$ \\
		\hline
		Coherent & $0.03614(1{\pm}0.20\%)$ & $\boldsymbol\times$ & $\boldsymbol\times$ & $\boldsymbol\times$ & $-4.8{\cdot}10^{-5}(1{\pm}43\%)$ \\
		\hline
		TMSV$_1$ & $0.03801(1{\pm}0.19\%)$ & $\boldsymbol\times$ & $\boldsymbol\checkmark$ & $\boldsymbol\times$ & $-3.7{\cdot}10^{-3}(1{\pm}37\%)$ \\
		TMSV$_2$ & $0.10582(1{\pm}0.080\%)$ & $\boldsymbol\times$ & $\boldsymbol\checkmark$ & $\boldsymbol\times$ & $-5.0{\cdot}10^{-3}(1{\pm}68\%)$ \\
		\hline
		SP$_1$ & $0.03768(1{\pm}0.28\%)$ & $\boldsymbol\times$ & $\boldsymbol\times$ & $\boldsymbol\checkmark$ & $-4.46{\cdot}10^{-5}(1{\pm}3.5\%)$ \\
		SP$_2$ & $0.07028(1{\pm}0.29\%)$ & $\boldsymbol\checkmark$ & $\boldsymbol\times$ & $\boldsymbol\checkmark$ & $-1.67{\cdot}10^{-4}(1{\pm}2.4\%)$ \\
		SP$_3$ & $0.09019(1{\pm}0.23\%)$ & $\boldsymbol\checkmark$ & $\boldsymbol\times$ & $\boldsymbol\checkmark$ & $-2.55{\cdot}10^{-4}(1{\pm}1.7\%)$ \\
		\hline\hline
	\end{tabular}
	\end{table*}

	The {\it higher-order conditional nonclassicality-number} $\mathfrak N_{c(b|a)}$ in Eq.~\eqref{eq:NclHO} is given in Table~\ref{tab:HONcl} together with a benchmark of the implemented methods.
	The classical coherent state and the TMSV state do not exhibit significant negativities and, thus, do not violate condition~\eqref{eq:NclHO}.
	By contrast, all SP states are clearly distinct from the classical upper bound, even state SP$_1$ (which has the lowest summed click number, see Table ~\ref{tab:HONcl}).
	Note that additional results of our analysis can be found in Sec.~E in the Supplemental Material~\cite{SM}.
	While joint quantum correlations are typically studied, the conditional quantum correlations considered here directly characterize the success of the measurement-induced generation of nonclassical states of light with imperfect detectors.
	This also includes the generation of nonclassicality exhibited in higher orders.

\paragraph*{Conclusions.--}
	We described and implemented rigorous and straightforwardly applicable approaches to uncovering quantum correlated light fields.
	We established a correlation coefficient for conditional statistics for accessing conditional quantum correlations measured with informationally incomplete click-counting detectors.
	For accessing quantum correlations of the joint statistics, we derived the bounds for Pearson's correlation coefficient for classical light.
	Applying both techniques, we successfully characterized nonclassical photon correlations for the experimentally generated light fields.
	The corresponding criteria are solely based on the measured click statistics without any need for knowing or correcting for the quantum efficiency, the dark count rate, and the exact response function of our detection system.
	A generalization to higher order moments of conditional statistics was also included.
	Conditional quantum correlations have been uncovered for split-photon states by using second- and higher-order moments criteria.
	The joint quantum correlations of two-mode squeezed-vacuum states have been identified via Pearson's correlation coefficient.

	In addition, our analysis is flexible in that it has straightforward extensions to general detection schemes based on click counting.
	In particular, this includes the cases of bright squeezed vacuum sources or when correlations occur in the temporal-spectral degree of freedom.
	Hence, our methods provide simple and yet powerful approaches for verifying different types of quantum correlated light fields for applications under realistic conditions.

\paragraph*{Acknowledgements.--}
	This work has received funding from the European Union Horizon 2020 Research and Innovation Program (QCUMbER, Grant Agreement No. 665148).
	MB is supported by a Rita Levi-Montalcini fellowship of MIUR.
	AD is partly supported by the UK EPSRC (EP/K04057X/2) and the UK National Quantum Technologies Programme (EP/M01326X/1, EP/M013243/1).
	I. A. W. acknowledges EPSRC (Grant No. EP/K034480/1), ERC (Grant MOQUACINO), and the UK National Quantum Technologies Programme.



\begin{thebibliography}{999}
	\bibitem{HBT56}
		R. Hanbury Brown and R. Q. Twiss,
		Correlation between Photons in two Coherent Beams of Light,
		\href{http://dx.doi.org/10.1038/177027a0}{Nature (London) \textbf{177}, 27 (1956)}.
	\bibitem{TG65}
		U. M. Titulaer and R. J. Glauber,
		Correlation Functions for Coherent Fields,
		\href{http://dx.doi.org/10.1103/PhysRev.140.B676}{Phys. Rev. \textbf{140}, B676 (1965)}.
	\bibitem{PADLA10}
		W. N. Plick, P. M. Anisimov, J. P. Dowling, H. Lee, and G. S. Agarwal,
		Parity detection in quantum optical metrology without number-resolving detectors,
		\href{http://dx.doi.org/10.1088/1367-2630/12/11/113025}{New J. Phys. \textbf{12}, 113025 (2010)}.
	\bibitem{DZTDSW11}
		A. Datta, L. Zhang, N. Thomas-Peter, U. Dorner, B. J. Smith, and I. A. Walmsley,
		Quantum metrology with imperfect states and detectors,
		\href{http://dx.doi.org/10.1103/PhysRevA.83.063836}{Phys. Rev. A \textbf{83}, 063836 (2011)}.
	\bibitem{CGLM14}
		F. Caruso, V. Giovannetti, C. Lupo, and S. Mancini,
		Quantum channels and memory effects,
		\href{http://dx.doi.org/10.1103/RevModPhys.86.1203}{Rev. Mod. Phys. \textbf{86}, 1203 (2014)}.
	\bibitem{KMNRDM07}
		P. Kok, W. J. Munro, K. Nemoto, T. C. Ralph, J. P. Dowling, and G. J. Milburn,
		Linear optical quantum computing with photonic qubits,
		\href{http://dx.doi.org/10.1103/RevModPhys.79.135}{Rev. Mod. Phys. \textbf{79}, 135 (2007)}.
	\bibitem{WJD07}
		H. M. Wiseman, S. J. Jones, and A. C. Doherty,
		Steering, Entanglement, Nonlocality, and the Einstein-Podolsky-Rosen Paradox,
		\href{http://dx.doi.org/10.1103/PhysRevLett.98.140402}{Phys. Rev. Lett. \textbf{98}, 140402 (2007)}.
	\bibitem{CPMA14}
		P. Chowdhury, T. Pramanik, A. S. Majumdar, and G. S. Agarwal,
		Einstein-Podolsky-Rosen steering using quantum correlations in non-Gaussian entangled states,
		\href{http://dx.doi.org/10.1103/PhysRevA.89.012104}{Phys. Rev. A \textbf{89}, 012104 (2014)}.
	\bibitem{KDM77}
		H. J. Kimble, M. Dagenais, and L. Mandel,
		Photon Antibunching in Resonance Fluorescence,
		\href{http://dx.doi.org/10.1103/PhysRevLett.39.691}{Phys. Rev. Lett. \textbf{39}, 691 (1977)}.
	\bibitem{M79}
		L. Mandel,
		Sub-Poissonian photon statistics in resonance fluorescence,
		\href{http://dx.doi.org/10.1364/OL.4.000205}{Opt. Lett. \textbf{4}, 205 (1979)}.
	\bibitem{AT92}
		G. S. Agarwal and K. Tara,
		Nonclassical character of states exhibiting no squeezing or sub-Poissonian statistics,
		\href{http://dx.doi.org/10.1103/PhysRevA.46.485}{Phys. Rev. A \textbf{46}, 485 (1992)}.
	\bibitem{V08}
		W. Vogel,
		Nonclassical Correlation Properties of Radiation Fields,
		\href{http://dx.doi.org/10.1103/PhysRevLett.100.013605}{Phys. Rev. Lett. \textbf{100}, 013605 (2008)}.
	\bibitem{MBWLN10}
		A. Miranowicz, M. Bartkowiak, X. Wang, Yu-xi Liu, and F. Nori,
		Testing nonclassicality in multimode fields: A unified derivation of classical inequalities,
		\href{http://dx.doi.org/10.1103/PhysRevA.82.013824}{Phys. Rev. A \textbf{82}, 013824 (2010)}.
	\bibitem{ALCS10}
		M. Avenhaus, K. Laiho, M. V. Chekhova, and C. Silberhorn,
		Accessing Higher Order Correlations in Quantum Optical States by Time Multiplexing,
		\href{http://dx.doi.org/10.1103/PhysRevLett.104.063602}{Phys. Rev. Lett. \textbf{104}, 063602 (2010)}.
	\bibitem{AOB12}
		A. Allevi, S. Olivares, and M. Bondani,
		Measuring high-order photon-number correlations in experiments with multimode pulsed quantum states,
		\href{http://dx.doi.org/10.1103/PhysRevA.85.063835}{Phys. Rev. A \textbf{85}, 063835 (2012)}.
	\bibitem{BC10}
		G. S. Buller and R. J. Collins,
		Single-photon generation and detection,
		\href{http://dx.doi.org/10.1088/0957-0233/21/1/012002}{Meas. Sci. Technol. \textbf{21}, 012002 (2010)}.
	\bibitem{ZABGGBRP05}
		G. Zambra, A. Andreoni, M. Bondani, M. Gramegna, M. Genovese, G. Brida, A. Rossi, and M. G. A. Paris,
		Experimental Reconstruction of Photon Statistics without Photon Counting,
		\href{http://dx.doi.org/10.1103/PhysRevLett.95.063602}{Phys. Rev. Lett. \textbf{95}, 063602 (2005)}.
	\bibitem{BGGMPTPOP11}
		G. Brida, M. Genovese, M. Gramegna, A. Meda, F. Piacentini, P. Traina, E. Predazzi, S. Olivares, and M. G. A. Paris,
		Quantum State Reconstruction Using Binary Data from On/Off Photodetection,
		\href{http://dx.doi.org/10.1166/asl.2011.1204}{Adv. Sci. Lett. \textbf{4}, 1 (2011)}.
	\bibitem{WDSBY04}
		E. Waks, E. Diamanti, B. C. Sanders, S. D. Bartlett, and Y. Yamamoto,
		Direct Observation of Nonclassical Photon Statistics in Parametric Down-Conversion,
		\href{http://dx.doi.org/10.1103/PhysRevLett.92.113602}{Phys. Rev. Lett. \textbf{92}, 113602 (2004)}.
	\bibitem{HPHP05}
		O. Haderka, J. Pe\v{r}ina, Jr., M. Hamar, and J. Pe\v{r}ina,
		Direct measurement and reconstruction of nonclassical features of twin beams generated in spontaneous parametric down-conversion,
		\href{http://dx.doi.org/10.1103/PhysRevA.71.033815}{Phys. Rev. A \textbf{71}, 033815 (2005)}.
	\bibitem{LBFD08}
		E. Lantz, J.-L. Blanchet, L. Furfaro, and F. Devaux,
		Multi-imaging and Bayesian estimation for photon counting with EMCCDs,
		\href{http://dx.doi.org/10.1111/j.1365-2966.2008.13200.x}{Mon. Not. R. Astron. Soc. \textbf{386}, 2262 (2008)}.
	\bibitem{BDFL08}
		J.-L. Blanchet, F. Devaux, L. Furfaro, and E. Lantz,
		Measurement of Sub-Shot-Noise Correlations of Spatial Fluctuations in the Photon-Counting Regime,
		\href{http://dx.doi.org/10.1103/PhysRevLett.101.233604}{Phys. Rev. Lett. \textbf{101}, 233604 (2008)}.
	\bibitem{CWB14}
		R. Chrapkiewicz, W. Wasilewski, and K. Banaszek,
		High-fidelity spatially resolved multiphoton counting for quantum imaging applications,
		\href{http://dx.doi.org/10.1364/OL.39.005090}{Opt. Lett. \textbf{39}, 5090 (2014)}.
	\bibitem{RHHPH03}
		J. \v{R}eh\`a\v{c}ek, Z. Hradil, O. Haderka, J. Pe\v{r}ina Jr., and M. Hamar,
		Multiple-photon resolving fiber-loop detector,
		\href{http://dx.doi.org/10.1103/PhysRevA.67.061801}{Phys. Rev. A \textbf{67}, 061801(R) (2003)}.
	\bibitem{ASSBW03}
		D. Achilles, C. Silberhorn, C. \'Sliwa, K. Banaszek, and I. A. Walmsley,
		Fiber-assisted detection with photon number resolution,
		\href{http://dx.doi.org/10.1364/OL.28.002387}{Opt. Lett. {\bf 28}, 2387 (2003)}.
	\bibitem{FJPF03}
		M. J. Fitch, B. C. Jacobs, T. B. Pittman, and J. D. Franson,
		Photon-number resolution using time-multiplexed single-photon detectors,
		\href{http://dx.doi.org/10.1103/PhysRevA.68.043814}{Phys. Rev. A \textbf{68}, 043814 (2003)}.
	\bibitem{LFCPSREPW08}
		J. S. Lundeen, A. Feito, H. Coldenstrodt-Ronge, K. L. Pregnell, Ch. Silberhorn, T. C. Ralph, J. Eisert, M. B. Plenio, and I. A. Walmsley,
		Tomography of quantum detectors,
		\href{http://dx.doi.org/10.1038/nphys1133}{Nature Phys. \textbf{5}, 27 (2008)}.
	\bibitem{SVA12}
		J. Sperling, W. Vogel, and G. S. Agarwal,
		True photocounting statistics of multiple on-off detectors,
		\href{http://dx.doi.org/10.1103/PhysRevA.85.023820}{Phys. Rev. A {\bf 85}, 023820 (2012)}.
	\bibitem{R14}
		R. Chrapkiewicz,
		Photon counts statistics of squeezed and multimode thermal states of light on multiplexed on-off detectors,
		\href{http://dx.doi.org/10.1364/JOSAB.31.0000B8}{J. Opt. Soc. Am. B \textbf{31}, B8 (2014)}.
	\bibitem{MSB16}
		F. M. Miatto, A. Safari, and R. W. Boyd,
		Theory of multiplexed photon number discrimination,
		\href{http://arxiv.org/abs/1601.05831}{arXiv:1601.05831 [quant-ph]}.
	\bibitem{SVA12a}
		J. Sperling, W. Vogel, and G. S. Agarwal,
		Sub-Binomial Light,
		\href{http://dx.doi.org/10.1103/PhysRevLett.109.093601}{Phys. Rev. Lett. {\bf 109}, 093601 (2012)}.
	\bibitem{BDJDBW13}
		T. J. Bartley, G. Donati, X.-M. Jin, A. Datta, M. Barbieri, and I. A. Walmsley,
		Direct Observation of Sub-Binomial Light,
		\href{http://dx.doi.org/10.1103/PhysRevLett.110.173602}{Phys. Rev. Lett. \textbf{110}, 173602 (2013)}.
	\bibitem{HSPGHNVS16}
		R. Heilmann, J. Sperling, A. Perez-Leija, M. Gr\"afe, M. Heinrich, S. Nolte, W. Vogel, A. Szameit,
		Harnessing click detectors for the genuine characterization of light states,
		\href{http://dx.doi.org/10.1038/srep19489}{Sci. Rep. \textbf{6}, 19489 (2016)}.
	\bibitem{SBVHBAS15}
		J. Sperling, M. Bohmann, W. Vogel, G. Harder, B. Brecht, V. Ansari, and C. Silberhorn,
		Uncovering Quantum Correlations with Time-Multiplexed Click Detection,
		\href{http://dx.doi.org/10.1103/PhysRevLett.115.023601}{Phys. Rev. Lett. {\bf 115}, 023601 (2015)}.
	\bibitem{SVA13}
		J. Sperling, W. Vogel, and G. S. Agarwal,
		Correlation measurements with on-off detectors,
		\href{http://dx.doi.org/10.1103/PhysRevA.88.043821}{Phys. Rev. A \textbf{88}, 043821 (2013)}.
	\bibitem{MLSWUSW08}
		P. J. Mosley, J. S. Lundeen, B. J. Smith, P. Wasylczyk, A. B. U'Ren, C. Silberhorn, and I. A. Walmsley,
		Heralded generation of ultrafast single photons in pure quantum states,
		\href{http://dx.doi.org/10.1103/PhysRevLett.100.133601}{Phys. Rev. Lett. {\bf 100}, 133601 (2008)}.
	\bibitem{SM}
		See Supplemental Material at \href{http://link.aps.org/supplemental/10.1103/PhysRevLett.117.083601}{this URL}, for additional aspect of the theory and the data analysis, which includes Refs.~\cite{SVA12,B69,M79,SBVHBAS15,SVA13,SVA12a}
	\bibitem{B69}
		D. R. Brillinger,
		The calculation of cumulants via conditioning,
		\href{http://dx.doi.org/10.1007/BF02532246}{Ann. Inst. Stat. Math. \textbf{21}, 215 (1969)}.
	\bibitem{P95}
		K. Pearson,
		Notes on regression and inheritance in the case of two parents,
		\href{http://dx.doi.org/10.1098/rspl.1895.0041}{Proc. R. Soc. Lond \textbf{58}, 240 (1895)}.
\end{thebibliography}
\end{document}